\documentclass[12pt]{iopart}

\newcommand{\width}{8.5cm}

\usepackage{graphicx}

\newcommand{\FIGPDF}[2]
{
       \typeout{FIGURE #1.pdf}
       \begin{figure}[h]
       \centering
       \includegraphics[width=\width]{#1.pdf}
       \caption{#2}
       \label{#1}
       \end{figure}
}

\begin{document}

\title{Random networks of carbon nanotubes optimized for transistor mass-production: multi-variable problem}

\author{M. \v{Z}e\v{z}elj}
\address{School of Electrical Engineering, University of Belgrade,
Bulevar kralja Aleksandra 73, 11120 Belgrade, Serbia}
\address{Scientific Computing Laboratory, Institute of Physics Belgrade, University of Belgrade, Pregrevica 118, 11080 Belgrade, Serbia}
\author{I. Stankovi\'{c}}
\address{Scientific Computing Laboratory, Institute of Physics Belgrade, University of Belgrade, Pregrevica 118, 11080 Belgrade, Serbia}
\ead{igor.stankovic@ipb.ac.rs}

\begin{abstract}
Random networks of single-walled carbon nanotubes (CNTs) usually contain both
metallic (m-CNTs) and semiconducting (s-CNTs) nanotubes with an approximate
ratio of 1:2, which leads to a trade-off between on-conductance and on/off
ratio. We demonstrate how the design problem can be solved with a realistic numerical
approach. We determine CNT density, length, and channel dimensions for which CNT thin-film transistors
(TFTs) simultaneously attain on-conductance higher than $1~{\rm \mu S}$
and on/off ratio higher than $10^4$. Fact that
asymmetric systems have more pronounced finite-size scaling behavior than symmetric,
enables us additional design freedom. A realisation probability of desired characteristics higher
than 99\% is obtained only with channel aspect length to width ratios $L_{\rm CH} / W_{\rm CH} < 1.2$
and normalized channel size $L_{\rm CH}W_{\rm CH} / l^2_{\rm CNT} > 250$ for a range of
CNT lengths $l_{\rm CNT} = 4-20~{\rm \mu m}$.
\end{abstract}



\maketitle

\section{Introduction}

Recently, random carbon nanotube networks have been demonstrated
as potential active materials in electronics applications \cite{2008_nature_Cao},
optoelectronics \cite{2013_science_Volder}, and sensors \cite{2014_nt_Abdelhalim}.
Carbon nanotube thin-film transistors are expected to enable
the fabrication of high-performance, flexible and transparent
devices using relatively simple techniques \cite{2008_acsn_Engel,
2011_nn_Sun, 2011_apl_Chandra, 2012_nl_Wang, 2013_ns_Park, 2013_nl_Lau,
2015_apl_Liu, 2015_arpc_Li}. As-grown networks of single-walled carbon nanotubes (CNTs) contain
both metallic (m-CNTs) and semiconducting (s-CNTs) nanotubes, which leads
to a trade-off between on-conductance and on/off conductance ratio
\cite{2006_apl_Kumar, 2007_nl_Kocabas, 2011_acsnano_Rouhi}.
If the density of CNTs in a thin-film transistor (TFT) is sufficiently high
that m-CNTs exceed the percolation threshold, the CNT network will
become predominantly metallic and, hence, the on/off ratio is very
small. In contrast, if the CNT density is so low that a conduction
path through the m-CNTs does not exist, a high on/off ratio
can be attained, but under such circumstances the low on-conductance
is disadvantage \cite{2010_apl_Sangwan}.

Various experimental efforts
have been made in order to increase both on-conductance and on/off
ratio of CNT TFTs. With roughly 1/3 as-grown CNTs being metallic,
extra steps such as, selective removal of m-CNTs via electrical
breakdown method \cite{2001_science_Collins}, are used in order to
cut the metallic paths through the transistors. However, such
breakdown also removes some s-CNT pathways leading to significant
decrease in the on-conductance. Other authors used semiconducting
enriched CNTs in order to enhance the performance of CNT TFTs. For
example, methods which separate CNTs by electronic type, such us
centrifugation of compositions of CNTs with surface active components
in density-gradient media \cite{2006_nnt_Arnold} or gas-phase plasma
hydrocarbonation reaction technique \cite{2006_science_Zhang} are
used during or after CNTs synthesis process. However, these techniques
also create defects in the remaining CNT networks and add impurities,
which degrade the overall performance of the TFTs
\cite{2011_jep_Kumar, 2010_cr_Hu, 2011_am_Hecht}. This adds more
difficulties to the fabrication process and the repeatability and
uniformity of devices are uncertain \cite{2011_jep_Kumar}.
The CNT TFTs are expected to be used in the drivers
and logic circuits of future devices, including low-cost,
printable e-papers, and radio-frequency identification
tags \cite{2014_sst_Che}. For such applications, it is
advantageous that the functional material for the active
layer be deposited on a flexible substrate using a fast,
room-temperature, and  non-vacuum processing.
The effects of m-CNTs in a random network can be reduced by
carefully controlling the CNT density, length and device geometry such
that the metallic fraction of the CNTs is bellow the percolation
threshold \cite{2008_nature_Cao}, i.e., every conducting path
contains at least one semiconducting CNT.

An optimized device,
i.e., the highest possible on-conductance at a given on/off
ratio has the total density of CNTs above the percolation
threshold and density of m-CNTs bellow the percolation
threshold. Although such high quality devices have been reported
in the literature \cite{2010_apl_Sangwan, 2003_apl_Snow, 2007_apl_Li,
2010_epl_Seppala}, numerical simulations and experiments to
determine the CNT density, channel size and CNT length for optimum
device performance fabricated at industrial yield rates are
lacking.

In this paper, we study the effect of the device
parameters (density of CNTs, channel dimensions and CNT length)
on electrical properties in order to obtain optimized and uniform
device performance without using any postgrowth treatment.
We identify probabilities of different conduction regimes of random CNT network based
on our recently determined scaling laws for asymmetric
systems of percolating sticks \cite{2012_pre_Zezelj}.
In this work, we demonstrate how geometrical aspects contribute to feasibility
random CNT networks as switches: good transistor performance (i.e., high on-current and
on/off ratio) and uniform performance of realized
devices.


\section{Numerical method}

Monte Carlo (MC) simulations are coupled with an efficient
iterative algorithm implemented on the grid platform and used to
investigate electrical properties of CNT
networks \cite{2012_pre_Zezelj, 2012_prb_Zezelj, 2002_cpc_Stankovic,
2012_cpc_Balaz}. Details on comparison of model with existing
experimental data found in literature can be found in Supplementary
Information. We consider the two-dimensional systems
with isotropically placed CNTs modeled as width-less sticks with
a fixed length $l_{\rm CNT}$. The centers of the CNTs are randomly
positioned and oriented inside a channel with length
$L_{\rm CH}$ and width $W_{\rm CH }$. Source and drain electrodes
are placed at the left and right sides of the channel.
The top and bottom  boundaries of the system are free and
nonconducting, because the free boundary conditions are
more consistent with the CNT networks in practice. The behavior
of the CNT network is studied in terms of the normalized CNT
density $n = N / \mathcal{L}^2$, where $N$ is the total number of
CNTs and $\mathcal{L} = \sqrt{L W}$ is the normalized channel size,
where $L = L_{\rm CH} / l_{\rm CNT}$ is the normalized channel length
and $W = W_{\rm CH} / l_{\rm CNT}$ is the normalized channel width,
see \cite{2011_nt_Heitz}. The aspect ratio of the
system $r$ is defined as the ratio of the channel length and
width $r = L / W$, see \cite{2012_pre_Zezelj}. Without
elaborated postgrowth treatments, CNTs synthesized using
any available method are heterogeneous in the sense that
they are always a mixture of metallic and semiconducting
nanotubes with an approximate ratio of 1:2~\cite{2000_science_Fuhrer,2010_cr_Hu}, i.e., the fraction
of the m-CNTs is $f_{\rm M} = 1 / 3$ and the rest are s-CNTs
with fraction $f_{\rm S} = 2 / 3$~\footnote{The CNTs exist as metallic or semiconducting, based on their
chirality and composition.}.

We only consider long-channel limits ($L_{\rm CH} > l_{\rm CNT}$)
consistent with macroelectronics \cite{2006_apl_Kumar} and
low-bias conditions under which nonlinear effects are
negligible \cite{2000_science_Fuhrer}. For long-channel limits
conduction in the CNT network is described by percolation
theory as being that of a non-classical two-dimensional
conductor \cite{2012_prb_Zezelj}. Two sticks (CNTs) belong to the same
cluster if they intersect. The system percolates (conducts) if the
electrodes (source and drain) are connected with the same cluster \cite{2003_ipt_Stauffer}.
\footnote{The percolation threshold of the infinite-size stick system
is defined by the critical density
$n_{\rm c} \approx 5.64$ \cite{2012_pre_Zezelj}.
Similarly, the percolation threshold of only s-CNTs
is defined by the critical density
$n_{\rm c} / f_{\rm S} \approx 8.46$ and the percolation
threshold of only m-CNTs is $n_{\rm c} / f_{\rm M} \approx 16.9$.}

The conductance along a CNT segment in on-state $G_{\rm seg}$
is assumed uniform and can be calculated using \cite{2007_apl_Li,
2010_epl_Seppala, 2007_prb_Behnam}
\begin{equation}
G_{\rm seg} = \frac{4 e^2}{h} \frac{\lambda}{\lambda + l_{\rm seg}}, \label{G}
\end{equation}
where $e$ is the electron charge, $h$ is the Planck's constant,
$l_{\rm seg}$ is the length of the CNT segment, and $\lambda$ is the mean
free path of the electrons that is taken as $1.0~{\rm \mu m}$ for
m-CNTs and $0.3~{\rm \mu m}$ for s-CNTs \cite{2004_prl_Javey,
2004_mrs_Avouris, 2007_prl_Purewal}. We consider that
the conductance of a m-CNT is independent of the gate
voltage \cite{2000_science_Fuhrer} and in off-state
is also given by equation (\ref{G}). At the same time,
we assume that the conductance of a s-CNT in off-state
is $10^{6}$ times lower than in on-state, since on/off
ratio is usually about $10^6$ for well-performing
transistors based on individual s-CNTs \cite{2001_science_Bachtold,
2003_nature_Javey}. Equation~(\ref{G}) assumes diffusive electrical
transport through the CNTs typical for the rodlike nanostructures
whose length is larger than the mean free path of the electrons
($l_{\rm CNT} \gg \lambda$). For the diffusive electrical
transport the electrical conductance of a stick segment
is inversely proportional to the length of the CNT segment
$(4e^2/h)(\lambda/l_{\rm seg})$, cf. \cite{2004_nl_Park, 2010_nn_Franklin}.
The conductance of the CNT whose length is lower than the mean free path of the
electrons ($l_{\rm CNT} \ll \lambda$) is near the ballistic transport
limit $4e^2/h$ and also can be assumed by equation (\ref{G}),
cf. \cite{2003_nature_Javey, 2004_prl_Javey}.

Internal nodes for contacts between pairs of CNTs are
distinguished from boundary nodes for contacts between CNTs and
the source/drain electrodes. The contact conductances for the
internal nodes are assigned the following values: (i) $0.1 e^2/h$
for the junction between two m-CNTs or between two s-CNTs and
(ii) 100 times lower conductance for the junction between one m-CNT
and one s-CNT, as we neglect the rectifying behavior under low-bias
conditions \cite{2000_science_Fuhrer, 2001_prb_Buldum}. The contact resistance
at the boundary nodes is neglected since electrodes fabricated using,
for example, Au \cite{2008_nature_Cao}, Pd \cite{2003_nature_Javey},
or  aligned arrays of m-CNTs \cite{2014_ns_Sarker} yield good
Ohmic contact to CNTs. Therefore, if a CNT intersects an electrode
the potential of the electrode is applied to the intersection point.
Kirchhoff's current law was used to balance the current flow through each
node of the created network. An iterative equation solver (i.e.,
conjugate gradient method with Jacobi preconditioner) has been
employed to solve a large system of the linear equations following
from the Kirchhoff's laws \cite{2012_prb_Zezelj}. After obtaining the
total source-drain current $I$ under an applied voltage $V$
the macroscopic electrical conductance of the system is evaluated
as $G = I / V$, see \footnote{On/off ratio is defined as a ratio of
on- and off-current, i.e., $I_{\rm ON} / I_{\rm OFF}$,
at the same source-drain voltage $V$ applied in on- and
off-state. We only consider low-bias conditions, i.e.,
$V \approx 0$, under which nonlinear effects are negligible
and on- and off-currents are given by
$I_{\rm ON} = G_{\rm ON} V$ and $I_{\rm OFF} = G_{\rm OFF} V$,
respectively. Therefore, the on/off current
ratio $I_{\rm ON} / I_{\rm OFF}$ is equal to the
on/off conductance ratio, i.e.,
$G_{\rm ON} / G_{\rm OFF}$.}. Finally, for
each set of system parameters, electrical conductances for on- and
off-state are calculated for more than $10^5$ independent MC
realizations for systems with normalized size $L = W = 2$
down to $10^4$ realizations for the largest system $L = W = 50$
studied. The results
obtained using our conductance model show an excellent agreement
with recently published experimental results, see Supplementary
Information.

\section{Symmetric-channel results}

The randomly generated CNT network, if conducting, belongs to one of three complementary CNT network regimes according
to their percolation characteristics: (i) neither m-CNT nor s-CNT
path exists but the whole network is percolated through a mixed
path comprising m- and s-CNTs ($\rm \overline{S} \hspace{0.25 mm}
\overline{M}$), (ii) only s-CNT paths exist ($\rm S
\overline{M}$), and (iii) at least one m-CNT path exists (M).

Figure~\ref{figure_1} illustrates the structure of the CNT
networks and the redistribution of the currents in on- and
off-state with increasing the CNT density $n$. When
the CNT density $n$ increases the randomly generated CNT network
moves from one operation regime to another. The first regime
($\rm \overline{S} \hspace{0.25 mm} \overline{M}$) corresponds
to situation when the density of CNTs ($n = 8$) is higher than
the percolation threshold of the entire network $n_{\rm c}$ but
lower than the percolation threshold of only s-CNTs
($n_{\rm c} < n < n_{\rm c} / f_{\rm S}$). In this situation
the percolation network consists mixed m- and s-CNTs. As a
result, the on-conductance $G_{\rm ON}$ and resulting
on-current $I_{\rm ON}$ are reduced by the presence
of the low-conductive s-CNT/m-CNT junctions, see
figure \ref{figure_1}(a). At the same time, since a m-CNT
path does not exist, off-conductance is low and,
therefore, on/off ratio is high, cf. figures \ref{figure_1}(a) and (b).
The second operation regime ($\rm S \overline{M}$) occurs for the
medium CNT density ($ n_{\rm c} / f_{\rm S} < n < n_{\rm c} / f_{\rm M}$)
when only s-CNT paths exist, and the current flows through
high-conductance s-CNT/s-CNT junctions resulting in high on-current
$I_{\rm ON}$, see figure \ref{figure_1}(c). However,
CNT density ($n = 12$) is still lower than the
percolation threshold of only m-CNTs and the CNT TFT
in off-state is not short-circuited through a m-CNT path.
Therefore, CNT network can simultaneously achieve high
on-conductance $G_{\rm ON}$ and high on/off ratio
$G_{\rm ON}/G_{\rm OFF}$, cf. figures \ref{figure_1}(c)
and \ref{figure_1}(d). Finally, the third regime ($\rm M$)
of CNT network according to the percolation characteristics
occurs at densities close and above the percolation threshold
of only m-CNTs, i.e., $n_{\rm c} / f_{\rm M}$. For high CNT density
($n = 16$) at least one m-CNT path exists and high on-conductance
is obtained, see Fig~\ref{figure_1}(e). On the other hand, in
off-state the CNT network is shorted through m-CNTs and,
hence, the on/off ratio is very low, cf. figures \ref{figure_1}(e)
and \ref{figure_1}(f). Therefore, CNT network in $\rm M$ operation
regime can not be used as an active media for transistors with a high
switching performance.

\FIGPDF{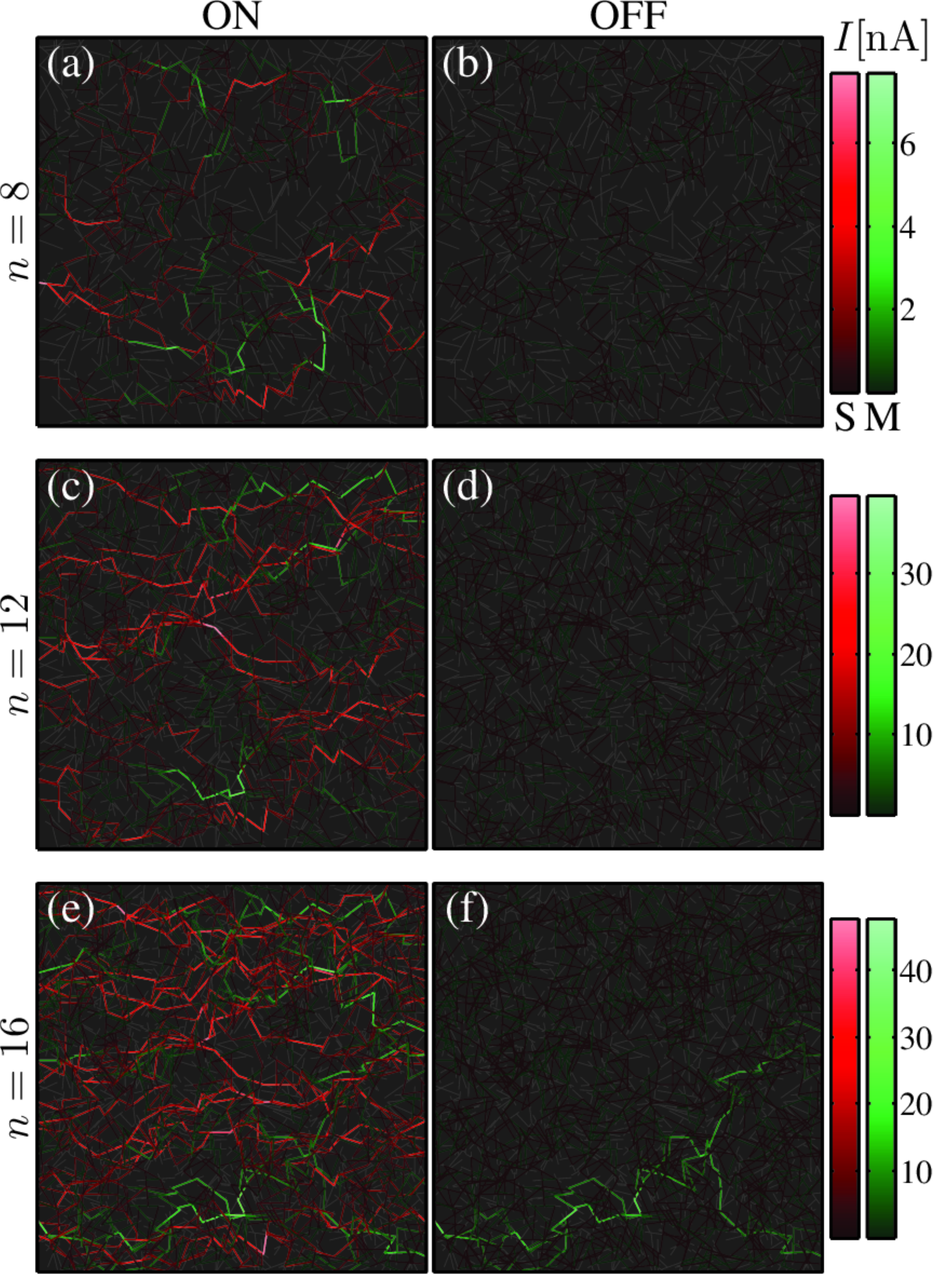}{Simulated on-
(a), (c), and (e) and off-current (b), (d), and (f) distribution for
different CNT densities. The source-drain voltage is
$V = 0.1~{\rm V}$, CNT length is $l_{\rm CNT} = 5~{\rm \mu m}$, and
channel dimensions are $L_{\rm CH}=W_{\rm CH} = 50~{\rm \mu m}$,
i.e., the normalized system size is $L = 10$. (a) When the density of
CNTs ($n = 8$) is significantly lower than the percolation
threshold of only s-CNTs ($n < n_{\rm c} / f_{\rm S}$) the
dominant percolation domain is through mixed paths comprising m-
and s-CNTs ($\rm \overline{S} \hspace{0.25 mm} \overline{M}$). (b)
Since there is no percolating path through only m-CNTs the off-current
is low, i.e., the on/off ratio is high. (c) For higher CNT
density ($n = 12$) only s-CNT paths exist ($\rm S \overline{M}$)
resulting in a high on-current $I_{\rm ON}$. (d) CNT
density is still lower than the percolation threshold of only
m-CNTs and therefore, the CNT TFT is not short-circuited in
off-state and on/off ratio is still high. (e) For high
CNT density ($n = 16$) m-CNT network percolates. (f) The CNT network in off-state is shorted through the m-CNTs.}

\FIGPDF{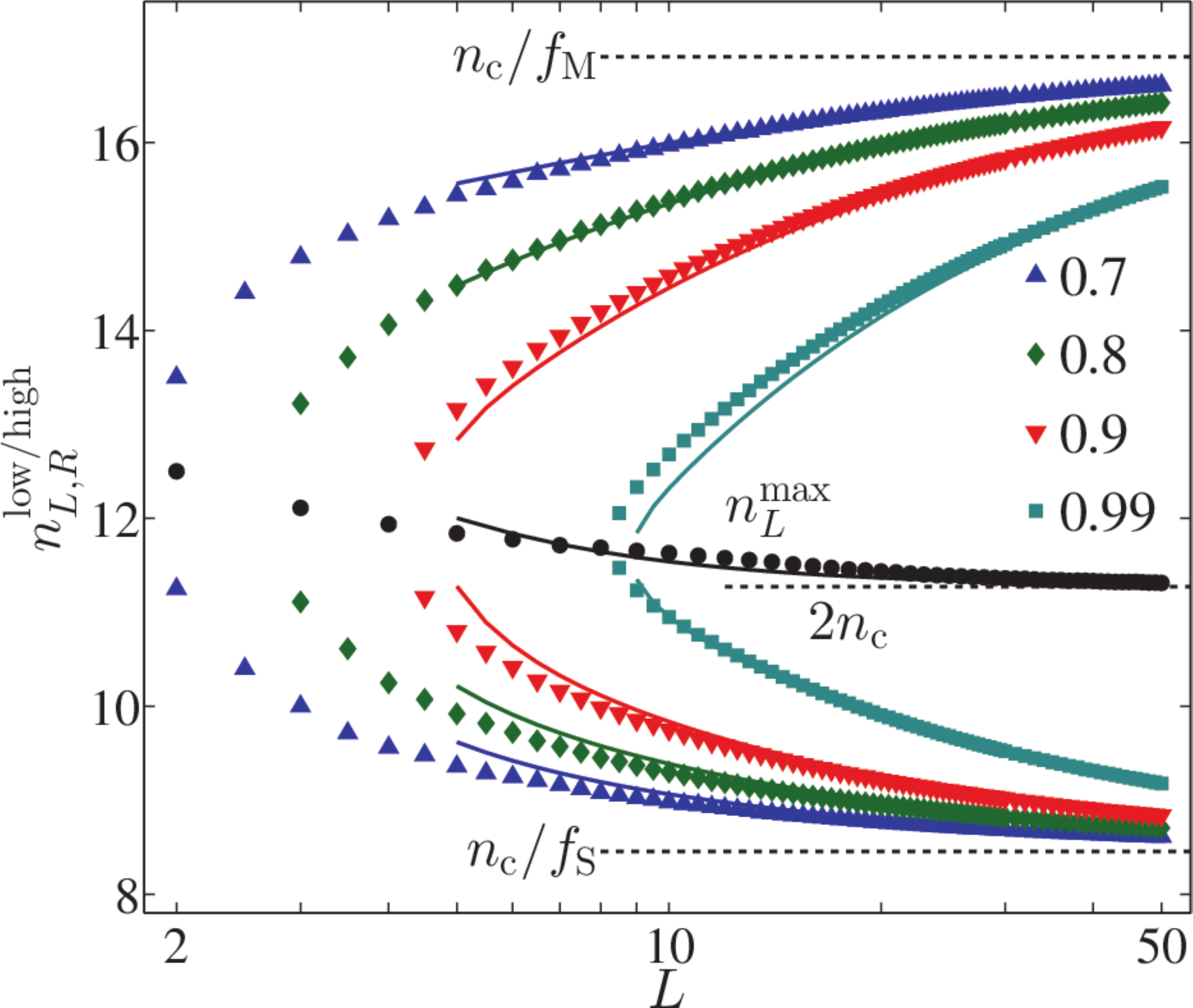}{The dependence of
$\rm S \overline{M}$-dominant region on the normalized
system size $L$, for different tolerances $R=0.7, 0.8, 0.9,$
and $0.99$. The points are the MC simulation results and
the solid lines are obtained using analytic model (see Supplementary Information).
With increasing the system size $L$, value of CNT density
$n_L^{\rm max}$, where percolation probability only through s-CNT paths
$R^{\rm S \overline{M}}_{n, L}$ reaches maximum, converges to
$2 n_{\rm c}$, low-bound density $n_{L, R}^{\rm low}$ to $n_{\rm c} /
f_{\rm S}$, and high-bound density $n_{L, R}^{\rm  high}$ to
$n_{\rm c} / f_{\rm M}$.}


We will determine region of CNT density $n$ where s-CNT paths are dominant($\rm S \overline{M}$) based on our recently determined parameters for moments of percolation probability distribution. In this region on-current and on/off ratio are expected to be high \footnote{Details of analytic model are given in Supplementary Information.}. We define a density range with lower $n^{\rm low}_{L, R}$ and upper boundary $n^{\rm high}_{L, R}$,  where probability of percolation only through s-CNT paths ($\rm S \overline{M}$) is higher then $R$ ($0\leq R\leq1$).
As one would expect, the range defined by $n^{\rm high}_{L, R}$ and
$n^{\rm low}_{L, R}$ increases with the normalized system size
$L$ and decreases with $R$, cf. figure \ref{figure_4}.
We can further observe interesting features of the shape of the
$\rm S \overline{M}$ percolation density range with increasing
the system size: (i) the position of $\rm S \overline{M}$
percolation probability maximum $n_L^{\rm max}$ converges to
$2 n_{\rm c}$ with increasing the system size,
(ii) low-bound density $n_{L, R}^{\rm low}$ decreases
and converges to the percolation threshold of
s-CNT paths $n_{\rm c} / f_{\rm S}$, and (iii) high-bound
density $n_{L, R}^{\rm high}$ increases and converges to
the percolation threshold of m-CNT paths
$n_{\rm c} / f_{\rm M}$, see figure \ref{figure_4}.

In the rest of this section,  we will quantify the dependence of calculated
on-conductance and on/off ratio on the CNT density, length,
and system size. Our aim is further to restrict parameter
range for which acceptable transistor characteristics are obtained with
a realisation probability higher than $99\%$.
For most kinds of integrated circuit applications acceptable
values for on-conductance and on/off ratio are
$1~{\rm \mu S}$ and $10^4$, respectively \cite{2010_cr_Hu,
2004_mrs_Avouris, 2010_acsnano_Wang}.



The results are shown in figure \ref{figure_5i} as a function of the CNT normalized
density $n$ for $l_{\rm CNT} = 5~{\rm \mu m}$ and the
system size $W_{\rm CH} = L_{\rm CH} = 100~{\rm \mu m}$, i.e.,
the normalized system size $L = 20$. The on-conductance
$G_{\rm ON}$ increases with CNT density $n$ and the 1st
and 99th percentiles converge to the median value, see
figure \ref{figure_5i}(a). The difference between the 99th and
1st percentile of the on/off ratio $G_{\rm ON} / G_{\rm OFF}$
reaches minimum close to $n^{\rm max}_{20}$ and rapidly increases
when the CNT density $n$ becomes higher than $n^{\rm high}_{20, 0.99}$,
see figure \ref{figure_5i}(b). For densities higher than
$n^{\rm high}_{20, 0.99}$ the percolation probability of
$\rm M$ configurations $R^{\rm M}_{n, L}$ is higher than 1\%,
i.e., $R^{\rm M}_{n, L} > 1 \%$ and therefore, the
1st percentile realization belongs to the $\rm M$ regime
and the sharp decrease in on/off behavior is obtained.
One can observe that maximum of on/off ratio is between the
densities $n^{\rm low}_{20, 0.99}$ and $n^{\rm high}_{20, 0.99}$.
More than 99\% of realized devices exhibit simultaneously on-conductance
higher than $1~{\rm \mu S}$ and on/off ratio higher than $10^4$
when the normalized density $n$ is close to $n^{\rm max}_{20}$,
cf. figures \ref{figure_5i}(a) and \ref{figure_5i}(b).

\FIGPDF{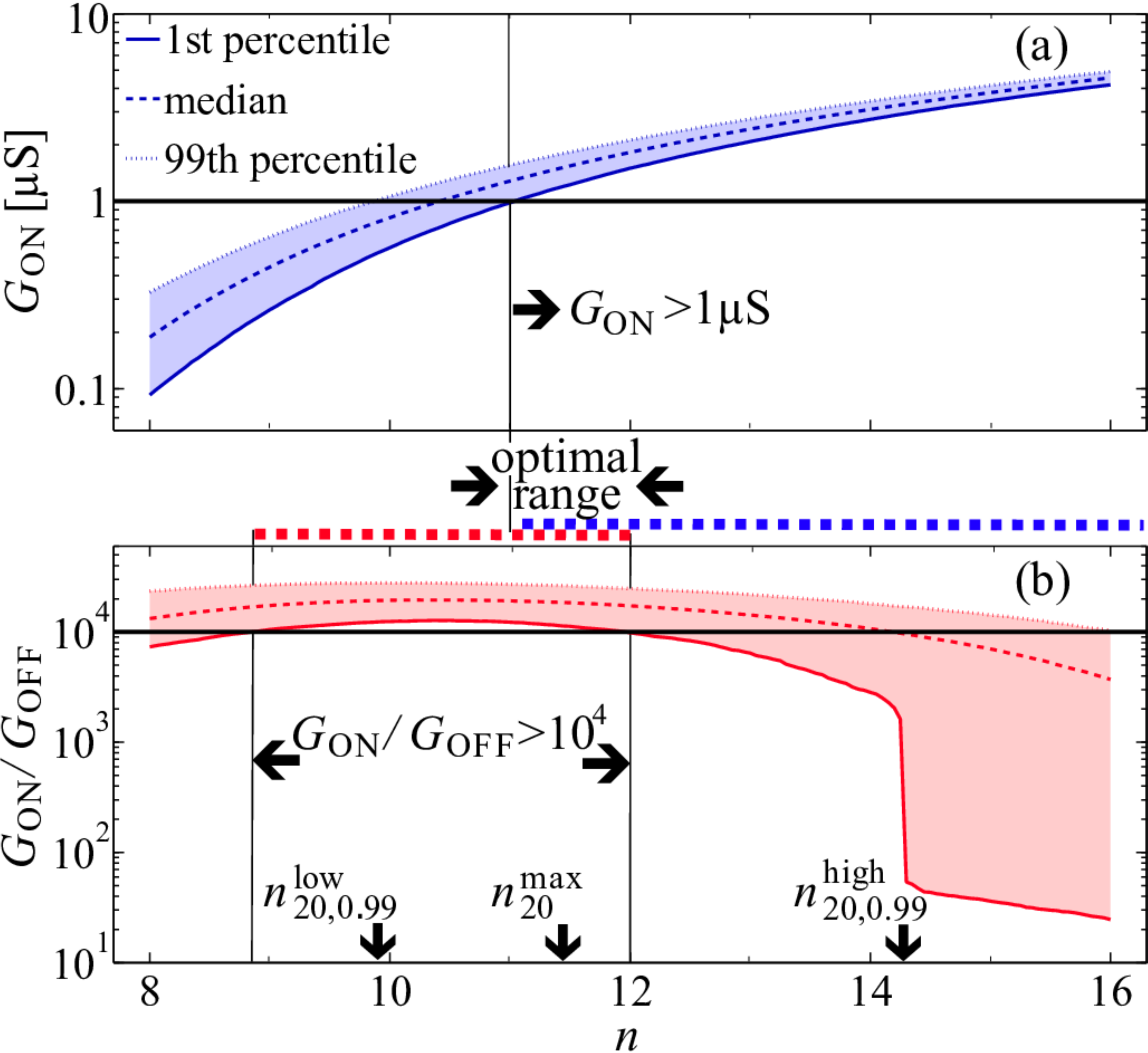}{On-conductance $G_{\rm ON}$
(a) and on/off ratio $G_{\rm ON} / G_{\rm OFF}$ (b) as a function
of the normalized CNT density $n$ for symmetric TFT with
the CNT length $l_{\rm CNT} = 5~{\rm \mu m}$ and channel
size $L_{\rm CH} = W_{\rm CH} = 100~{\rm \mu m}$, i.e., the
normalized system size $L = 20$. Solid line represents the
1st percentile, while the dashed and dotted lines correspond
to the median and 99th percentile, respectively, of the device
population. The arrows denote regions where more than 99\% of
devices have: (a) on-conductance higher than $1~{\rm \mu S}$
(horizontal bold line) or (b) on/off ratio higher than
$10^4$ (horizontal bold line). The positions of
lower $n^{\rm low}_{20, 0.99}$ and higher $n^{\rm high}_{20, 0.99}$
bounds of 0.99 percolation probability for $\rm S \overline{M}$
regime are also given, as well as position of $\rm S \overline{M}$
probability maximum $n^{\rm max}_{8, 50}$.}

A similar behavior of the on-conductance $G_{\rm ON}$ and
the on/off ratio $G_{\rm ON} / G_{\rm OFF}$ can be observed
in figure \ref{figure_6i}(a) with increasing the normalized
system size $L$. While the percentiles of the on-conductance
$G_{\rm ON}$ experience continuous convergence to the infinite
system value, the on/off ratio exhibits a sharp transition
in the 1st percentile behavior at the normalized system size
$L \approx 7$. For the system size bellow $L < 7$ the
percolation probability of realizations through only m-CNTs
is higher than 1\%, i.e., $R^{\rm M}_{n, L} > 1 \%$ and therefore,
the 1st percentile realization is short-circuited with
$G_{\rm ON} / G_{\rm OFF} < 10$, cf. Supplementary Information for probability distribution functions of on/off ratio.
On the other hand, more than 99\% of realized devices exhibit
simultaneously on-conductance higher than $1~{\rm \mu S}$
and on/off ratio higher than $10^4$ when the normalized
system size is $L > 16$, see figure \ref{figure_6i}(a).

\FIGPDF{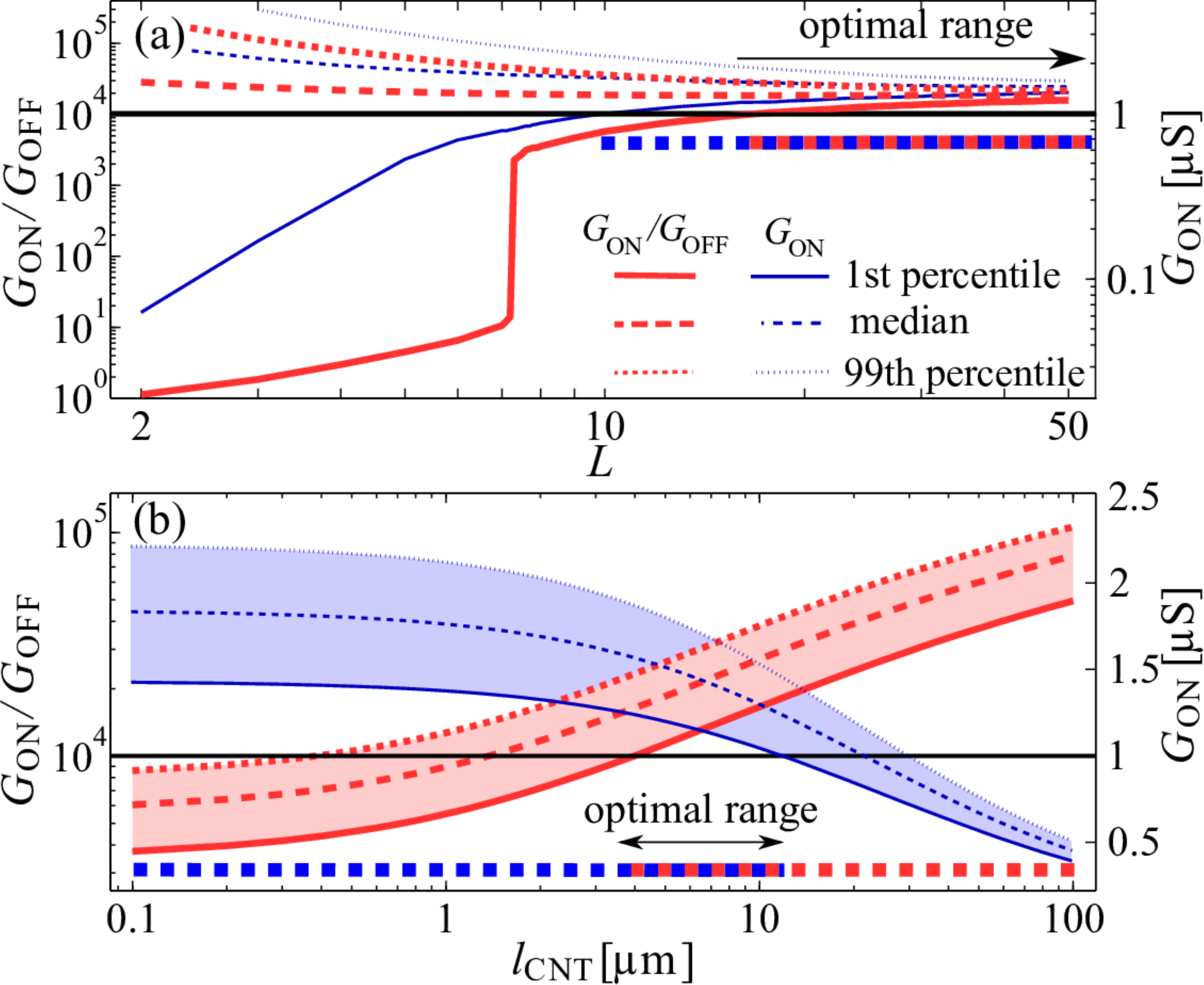}{The dependence of on-conductance
$G_{\rm ON}$ and on/off ratio $G_{\rm ON} / G_{\rm OFF}$
on (a) normalized system size $L$ for symmetric-channel TFTs
with CNT length $l_{\rm CNT} = 5~{\rm \mu m}$ and
normalized density $n^{\rm max}_L$ and (b) CNT length $l_{\rm CNT}$ for symmetric
system with the normalized size $L = 20$ and the normalized CNT density
$n$ equal to $n^{\rm max}_{20} = 11.45$. Solid line represents the
1st percentile, while the dashed and dotted lines
correspond to the median and 99th percentile, respectively,
of the device population. The arrows denote regions
where more than 99\% of devices have on-conductance higher
than $1~{\rm \mu S}$ and on/off
ratio higher than $10^4$ (horizontal bold line).}

The influence of CNT length $l_{\rm CNT}$ on the transistor
performance is explored in figure \ref{figure_6i}(b). Here,
the normalized system size is fixed and large $L = 20$,
in order to minimize influence of finite-size scaling
effects on the transistor performance, see also ~\ref{2015_carbon_Kuwahara}. Therefore,
on-conductance $G_{\rm ON}$ and on/off ratio
$G_{\rm ON} / G_{\rm OFF}$ depend only on the electrical
characteristics of the nanotubes, i.e., their length
$l_{\rm CNT}$, in accordance with equation (\ref{G}).
When the CNT length $l_{\rm CNT}$ is larger than the electron mean free path,
i.e., $l_{\rm CNT} > 1~{\rm \mu m}$, diffusive transport in CNTs
becomes dominant and the on-conductance $G_{\rm ON}$ of
the network starts to decrease linearly with the CNT length,
see figure \ref{figure_6i}(b). At the same time, due to a increased
resistance of m-CNTs, leak-currents in off-state through
the m-CNTs decrease and on/off ratio improves
with increasing the CNT length, see figure \ref{figure_6i}(b).
It is important to note that in this trade-off, the
improvement of on/off ratio is one order of magnitude
with increasing $l_{\rm CNT}$ from $1~{\rm \mu m}$
to $10~{\rm \mu m}$ while $G_{\rm ON}$ decreases only $10\%$.
However, more than 99\% of realized devices exhibit simultaneously
high on-conductance and high on/off ratio when the CNT
length is between $l_{\rm CNT} = 4 - 12~{\rm \mu m}$,
cf. figures \ref{figure_6i}(b).

\section{Asymmetric-channel results}

For the case of the asymmetric channel the dependence
of the normalized density $n_{L, W}^{\rm max}$ where
$\rm S \overline{M}$ percolation probability
reaches maximum on the normalized system
dimensions $L$ and $W$ is shown in Fig~\ref{figure_10}(a).
The normalized  density $n^{\rm max}_{L,W}$ increases with
increasing the normalized channel length $L$ and decreases
with increasing the width $W$, cf.
\ref{figure_10}(a) and Supplementary Information. The agreement between the MC simulation
results $n_{L, W}^{\rm max}$ and the values $n_{L, W}^{\rm model}$
obtained from our analytic model (cf. Supplementary Information)
is better than 9\% for the systems with $L > 5$,
see figure \ref{figure_10}(b). From figures \ref{figure_10}(c)
and \ref{figure_10}(d) we expect to achieve $99\%$ of only
s-CNTs conducting realizations roughly above the line
$L_{\rm CH} W_{\rm CH} > 250~l^2_{\rm CNT}$. Within that
region we see that $n^{\rm low}_{L,W,0.99}$ depends only
on the system width $W$. On the other hand, the upper limit
of $99\%$ confidence region, $n^{\rm high}_{L,W,0.99}$
depends on both the normalized length $L$ and width $W$.

\FIGPDF{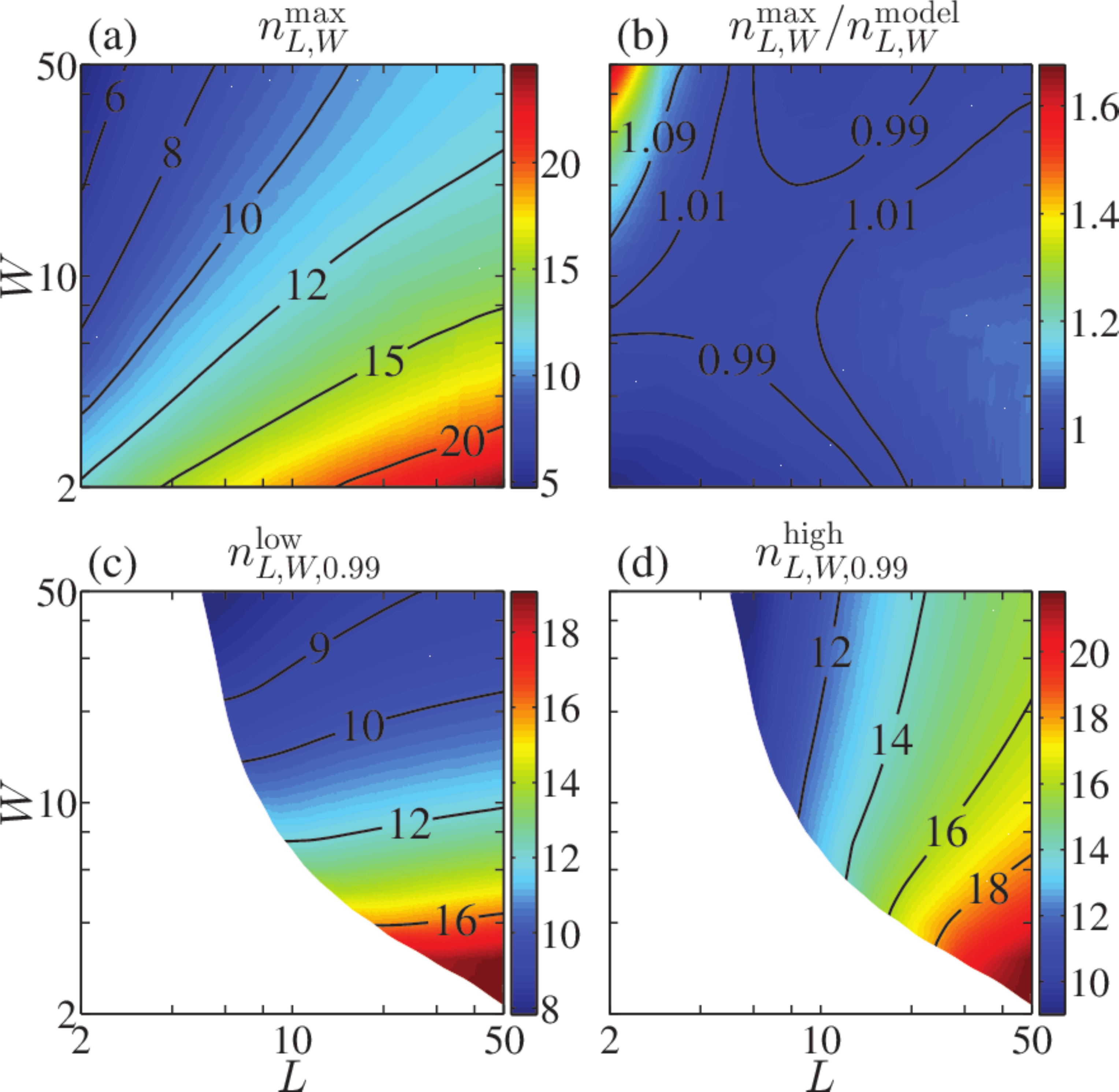}{The dependence
of (a) the normalized density $n_{L, W}^{\rm max}$
where the probability of $\rm S \overline{M}$ percolation
reaches maximum and (b) the ratio between the
MC simulation results $n_{L, W}^{\rm max}$ and
the values $n_{L, W}^{\rm model}$ are obtained using analytic model (cf., Supplementary Information) on the normalized
channel width $W$ and length $L$ . The low-bound
$n^{\rm low}_{L, W, 0.99}$ (c) and high-bound density
$n^{\rm high}_{L, W, 0.99}$ (d) determine a density range
where the probability of $\rm S \overline{M}$
realizations is higher than 0.99.}

The random CNT TFTs with an asymmetric channel have
similar characteristics of the on-conductance $G_{\rm ON}$
and on/off ratio $G_{\rm ON} / G_{\rm OFF}$ compared
to the symmetric-channel configurations, cf. figures \ref{figure_5i}
and \ref{figure_11i}. The on-conductance $G_{\rm ON}$ also
increases with increasing the CNT density $n$ and the 1st and 99th
percentiles converge to the median value, see figure
\ref{figure_11i}(a). The difference between the 99th and
1st percentile of on/off ratio $G_{\rm ON} / G_{\rm OFF}$
also reaches minimum close to the density where probability
of $\rm S \overline{M}$ percolation reaches maximum,
i.e., $n^{\rm max}_{8, 50}$, and rapidly increases when the
CNT density becomes higher than $n^{\rm high}_{8, 50, 0.99}$,
see figure \ref{figure_11i}(b). Similarly, the maximum of
the on/off ratio is between densities $n^{\rm low}_{8, 50, 0.99}$
and $n^{\rm high}_{8, 50, 0.99}$, and more than 99\% of realized
devices exhibit simultaneously on-conductance higher than
$1~{\rm \mu S}$ and on/off ratio higher than $10^4$ when
the normalized density $n$ is close to $n^{\rm max}_{8, 50}$,
cf. figures \ref{figure_11i}(a) and \ref{figure_11i}(b).
Hence, the density $n^{\rm max}_{L, W}$ for an asymmetric,
as well as symmetric, configurations can be used as a
compromise value for obtaining the optimized transistor
performance, see figures \ref{figure_5i} and \ref{figure_11i}.
However, the narrow-channel configurations generally have
a higher on-conductance compared to the symmetric channels (see Supplementary Information). Indeed, as can be seen
from figures \ref{figure_5i}(a) and \ref{figure_11i}(a),
narrow-channel TFT with the aspect ratio $r = 8/50$ has roughly
$1 /r \approx 6$ times higher on-conductance $G_{\rm ON}$ compared to the
symmetric-channel configuration with the same normalized channel size
$\mathcal{L} = 20$ and CNT length $l_{\rm CNT} = 5~{\rm \mu m}$.

\FIGPDF{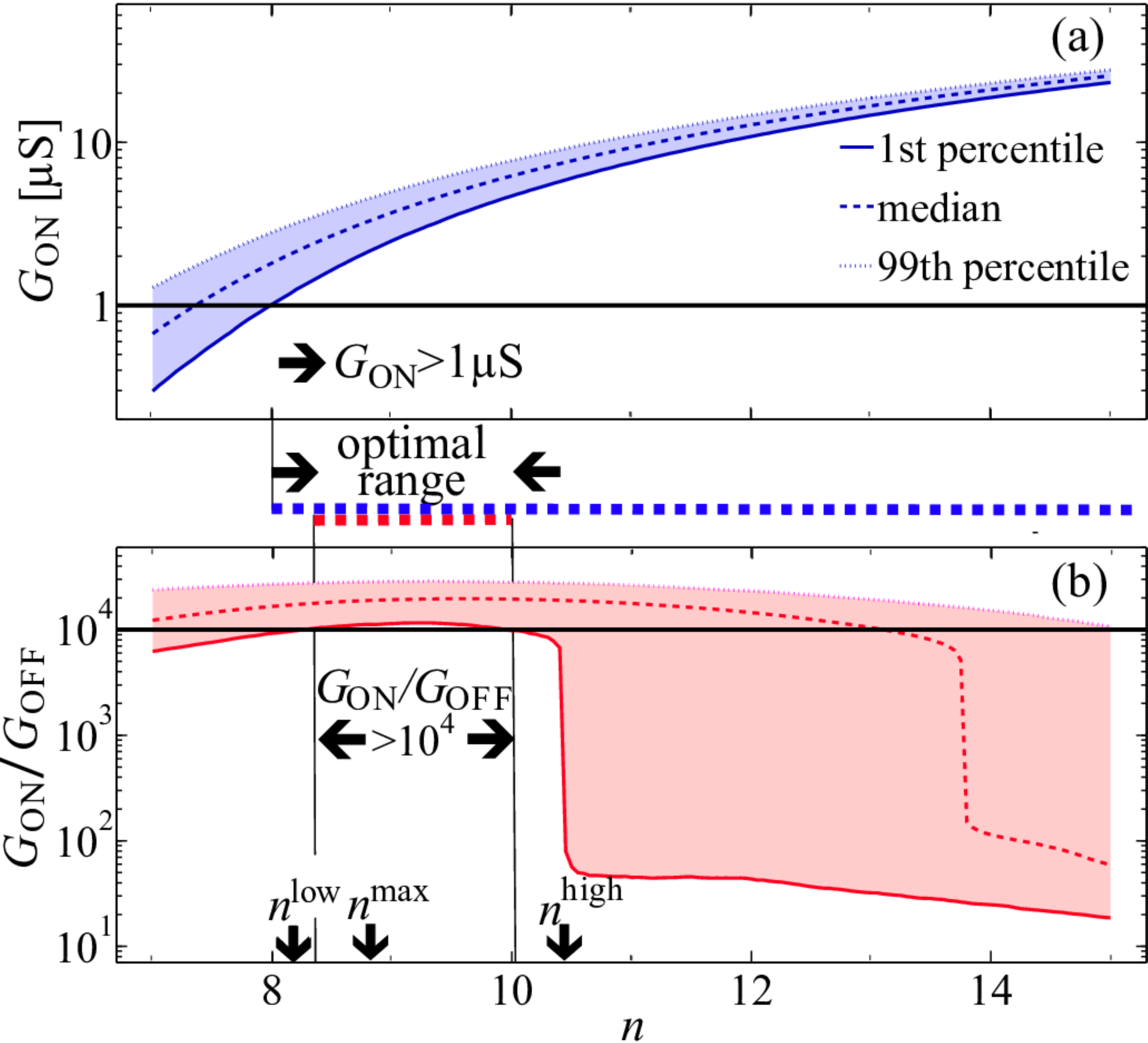}{On-conductance $G_{\rm ON}$
(a) and on/off ratio $G_{\rm ON} / G_{\rm OFF}$ (b) as
a function of the normalized CNT density $n$ for narrow-channel
TFT with $l_{\rm CNT} = 5~{\rm \mu m}$ and the channel
dimensions $L_{\rm CH} = 40~{\rm \mu m}$ and $W_{\rm CH} =
250~{\rm \mu m}$, i.e., the normalized system dimensions
$L = 8$ and $W = 50$. Solid line represents the 1st percentile,
while the dashed and dotted lines correspond to the median and
99th percentile, respectively, of the device population.
The arrows denote regions where more than 99\% of devices have:
(a) on-conductance higher than $1~{\rm \mu S}$ (horizontal bold line)
or (b) on/off ratio higher than $10^4$ (horizontal bold line).
The positions of lower $n^{\rm low}_{8, 50, 0.99}$ and higher
$n^{\rm high}_{8, 50, 0.99}$  bounds of 0.99 percolation
probability for $\rm S \overline{M}$ regime are also given,
as well as position of $\rm S \overline{M}$
probability maximum $n^{\rm max}_{8, 50}$.}

The 1st percentiles of the on-conductance $G_{\rm ON}$
and on/off ratio $G_{\rm ON} / G_{\rm OFF}$ are calculated
for the same structural parameters $L$, $W$, and
$n^{\rm max}_{L, W}$, and different CNT lengths
$l_{\rm CNT} = 4, 8, 12~{\rm \mu m}$, see figure \ref{figure_12}.
In accordance to the results shown for the symmetric-channel
configuration in figure \ref{figure_6i}(b) we note that with
increasing the CNT length $l_{\rm CNT}$ the on-conductance
$G_{\rm ON}$ decreases, while the on/off ratio
$G_{\rm ON} / G_{\rm OFF}$ increases, see figure \ref{figure_12}.
When the CNT length $l_{\rm CNT}$ is bellow $4~{\rm \mu m}$ ballistic
electrical transport becomes dominant and the on/off ratio
becomes lower than $10^4$, cf. figures \ref{figure_12}(b),
\ref{figure_12}(d), and \ref{figure_12}(f). On the other hand,
when $l_{\rm CNT}$ is higher than $20~{\rm \mu m}$ a high
on-conductance can be attained only for a very
large channel with dimensions of the order of $1~{\rm mm}$, see
figure \ref{figure_12}(e). Hence, the CNT length in the range
$l_{\rm CNT} = 4-20~{\rm \mu m}$ results in an acceptable
balance between the on-conductance and on/off performance of
the CNT TFTs.

Regions A and B in figure \ref{figure_12} represent the CNT
networks with a low 1st-percentile value of the on-conductance
$G_{\rm ON}$. The region A is defined as a region where
the probability of $\rm \overline{S} \hspace{0.25 mm} \overline{M}$
percolation is higher than 1\%, i.e.,
$R^{\rm \overline{S} \hspace{0.25 mm} \overline{M}}_{n, L, W} > 1\%$
and therefore, the 1st-percentile realization is mixed
path with low-conductive s-CNT/m-CNT junctions.
The realizations in the region B have a high value of the
aspect ratio $r$ and therefore, a low overall
on-conductance (see Supplementary Information). On the other hand,
regions C and D in figure \ref{figure_12} represent the CNT networks
with a low 1st-percentile value  of the on/off ratio
$G_{\rm ON} / G_{\rm OFF}$. The TFTs in the region C feature
networks with the probability of percolation through only m-CNTs
higher than 1\%, i.e., $R^{\rm M}_{n, L, W} > 1\%$ and therefore,
the 1st-percentile realization is short-circuited having a very
low $G_{\rm ON} / G_{\rm OFF}$. The networks operating
in D region are not short-circuited but the on/off ratio
is low because of high leak-currents in off-state
through the high-density m-CNTs. We can conclude that
optimal dimensions of CNT TFT channel regarding to the
device switching performance are those outside
the regions A, B, C, and D shown in figure \ref{figure_12}.
The dotted lines approximate the region where CNT TFTs
attain high on-conductance and at the same time high
on/off ratio. Therefore, the random CNT TFTs with the channel
aspect ratio $L_{\rm CH} / W_{\rm CH}  < 1.2$ and the
normalized size $L_{\rm CH}W_{\rm CH} / l^2_{\rm CNT} > 250$
with a probability higher than 99\% exhibit on-conductance higher than
$1~{\rm \mu S}$ and at the same time on/off ratio higher
than $10^4$.

\FIGPDF{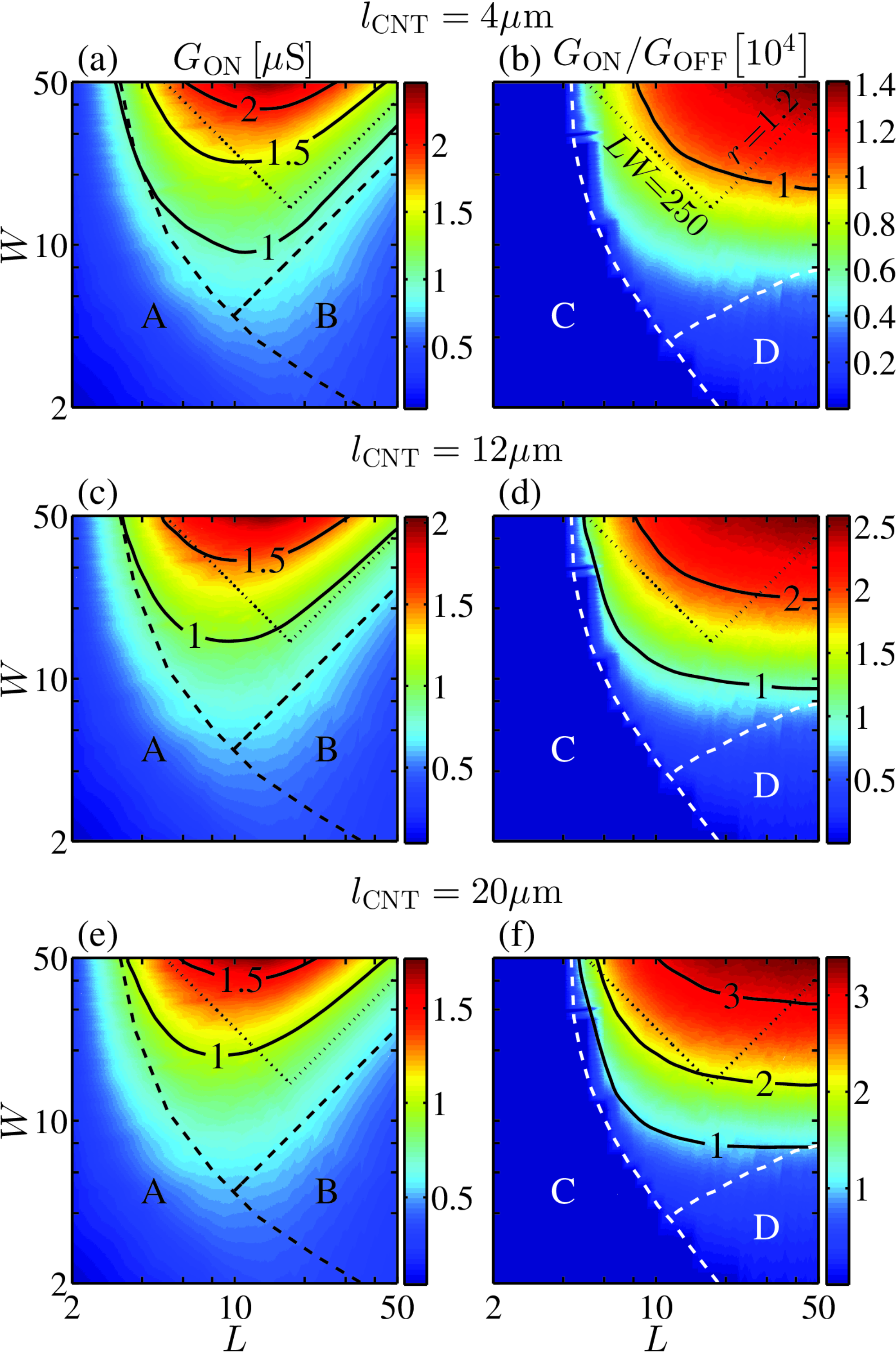}{The results for
the 1st percentile of the on-conductance $G_{\rm ON}$
(a), (c), and (e) and on/off ratio $G_{\rm ON} / G_{\rm OFF}$
(b), (d), and (f) calculated at $n_{L, W}^{\rm max}$ as
a function of the normalized channel width $W$ and
length $L$ for different CNT lengths
$l_{\rm CNT} = 4, 12, 20~{\rm \mu m}$. The region A
represents CNT TFT configurations with the probability of
$\rm \overline{S} \hspace{0.25 mm} \overline{M}$ percolation
higher than 1\%, i.e.,
$R^{\rm \overline{S} \hspace{0.25 mm} \overline{M}}_{n, L, W} > 1\%$,
while the region B represents configurations with a high aspect ratio
(here $r > 2$). TFTs in the C region feature networks
with the probability of $\rm M$ percolation
higher than 1\%, i.e., $R^{\rm M}_{n, L, W} > 1\%$, while
the region D denotes networks with a high CNT density
(here $n > 15$). The area above the dotted lines
($LW = 250$ and $L / W = 1.2$) can be used as an approximation
of the region where the random CNT TFTs simultaneously attain
high on-conductance and high on/off ratio.}

\section{Conclusions}

In this paper, we present the numerical simulation results
of switching performance of the transistors based on random networks
of as-grown CNTs. The CNT thin films studied here are viewed as a material for the
low-cost, flexible, and transparent field-effect transistors.
A factor that makes CNT films complex is that they contain both
metallic and semiconducting nanotubes. Only the s-CNTs have
highly modulated conductance by the gate and only junctions
between CNTs of the same type are highly conductive. Therefore,
the random CNT TFTs that percolate only through s-CNT paths
can simultaneously attain high on-conductance and high
on/off ratio. As a result, a key limitation in scaling-up
production of the random CNT TFTs with high on-current and
on/off ratio is a requirement to achieve uniform performance
of the realized devices.

We demonstrate utility of the percolation probability
functions for as-grown CNT networks.
The derived expressions
for the percolation probability functions represent an excellent
fit to the data calculated using simulations. We also show
that the fraction of 1:2 of metallic to semiconducting nanotubes
provides a sufficient design freedom. We avoid creating
m-CNT paths, while at the same time determine a density range
where the rate of realizations that percolate only through s-CNT paths
is higher than $99\%$. Since there is a trade-off between the on-conductance
and on/off ratio regarding to the CNT density we show that the position
of only s-CNT percolation probability maximum can be used
as a good balance for the CNT density value. We have included into
the CNT conductance model diffusive electrical transport through
the CNTs typical for the rodlike nanostructures whose
length is larger than the mean free path of the electrons.
This enabled us to consider influence of the CNT length on the
device characteristics. When the CNT length increases
diffusive transport in the CNTs becomes dominant and on-conductance
decreases with the CNT length. At the same time, because of
increased resistance of m-CNTs, the leak-currents in off-state
through the m-CNTs decrease and on/off ratio improves.
This results in a further trade off, still the improvement
of the on/off ratio is much larger than detrimental loss
of the on-conductance.

The asymmetric systems have smaller finite-size scaling exponents
than symmetric. This results in more pronounced finite-size
scaling behavior than in symmetric case and enables us even
more design freedom. We conclude that decrease of the
channel length is a good strategy for the performance
improvement for two reasons: (i) the on-conductance increases
with decreasing the aspect ratio and (ii) the percolation
only through s-CNT paths occurs at lower densities resulting in
the lower costs of TFT production. Following this, we have
performed a parameter study to find optimal channel dimensions
for CNT TFTs. We present a region of the channel dimension where
the most of the random CNT realizations have satisfactory
transistor performance. According to the criteria of a
high on-conductance and at the same time a high on/off ratio,
within $99\%$ confidence range, the optimal region of
channel dimensions can be estimated with aspect
ratio $L_{\rm CH} / W_{\rm CH} < 1.2$ and size
$L_{\rm CH}W_{\rm CH} / l^2_{\rm CNT} > 250$. This
conclusion remains valid when the CNT length belongs
to the range $l_{\rm CNT} = 4-20~{\rm \mu m}$ resulting
in an acceptable balance between the on-conductance and
on/off performance of the random CNT TFTs. At the end,
we have demonstrated that results of our model for
on-conductance and on/off ratio show an excellent
agreement with the recently published experimental
results. Hence, we conclude that the channel dimensions $L_{\rm CH}$
and $W_{\rm CH}$, CNT length $l_{\rm CNT}$, and density $n$
are the only parameters needed for description and optimization
of TFTs based on random networks of as-grown CNTs.

\ack
This work was supported by the Ministry of Education, Science,
and Technological Development of the Republic of Serbia under
project ON171017, OI1611005, and OI1611003, es well as,
by the European Commission under H2020 project DAFNEOX,
Grant No. 645658. Numerical simulations were
run on the PARADOX supercomputing facility at the Scientific
Computing Laboratory of the Institute of Physics Belgrade.


\end{document}